\documentclass[twocolumn,prl]{revtex4}
\usepackage{graphicx}% Include figure files
\usepackage{amsbsy}
\usepackage{color}
\usepackage[colorlinks,citecolor=blue]{hyperref}
\def\cp#1{\mathbf{#1}}

\begin{document}

\title{
Multichannel Molecular State and Rectified Short-range Boundary Condition for Spin-orbit Coupled Ultracold Fermions Near p-wave Resonances}
\author{Xiaoling Cui}
\affiliation{Beijing National Laboratory for Condensed Matter Physics, Institute of Physics, Chinese Academy of Sciences, Beijing 100190, China }
\date{{\small \today}}
\begin{abstract}
We study the interplay of spin-orbit coupling (SOC) and strong p-wave interaction to the scattering property of spin-1/2 ultracold Fermi gases. Based on a two-channel square-well potential generating p-wave resonance, we show that the presence of an isotropic SOC, even for its length much longer than the potential range, can greatly modify the p-wave short-range boundary condition(BC). As a result, the conventional p-wave BC cannot predict the induced molecules near p-wave resonance, which can be fully destroyed to vanish due to strong interference between s- and p-wave channels. By analyzing the intrinsic reasons for the breakdown of conventional BC, we propose a new p-wave BC that can excellently reproduce the exact molecule solutions and also equally apply for a wide class of single-particle potentials besides SOC. This work reveals the significant effect of SOC to both the short- and long-range properties of fermions near p-wave resonance, paving the way for future exploring interesting few- and many-body physics in such system.
\end{abstract}

\maketitle

The interplay of spin-orbit coupling (SOC) and interaction has generated tremendous research interests in condensed matter physics\cite{review_1,review_2}, while ultracold atomic gases offer an ideal platform for its study giving successful realizations of synthetic SOC using Raman lasers\cite{gauge2exp,fermisocexp1,fermisocexp2,2dsoczhang1,Wu2015} and highly tunable interactions via Feshbach resonances\cite{Chin}.  Nevertheless, before studying the complex many-body physics the very first question to address is how to model the fundamental two-body interactions. A crucial factor here is the asymptotic behavior of two-body wave function in the short-range limit, called the short-range boundary condition(BC), which is the basis for constructing the Huang-Yang pseudo-potentials\cite{LHY, Stock05} and also equivalent to the use of renormalized contact models\cite{review_Fermi_gas, Gurarie}. In the presence of SOC, studies have shown that the usual s-wave short-range BC, giving the conventional s-wave models, %in s-wave channel, giving the conventional s-wave model, 
is hardly modified near s-wave resonances given the typical length of realistic SOC  much longer than the potential range\cite{Cui, Yu, Zhang}. Despite the negligible short-range consequence, SOC can greatly change the long-range (low-energy) scattering properties from near the threshold\cite{Cui, Yu, Blume} to intermediate energy regime\cite{Blume,You, Greene}. Moreover, with conventional s-wave models it has been found that SOC can induce shallow molecules\cite{Vijay} and universal trimers\cite{trimer_soc, borromean_soc} more easily, and lead to various fascinating many-body phenomena in both bosons and fermions atomic systems\cite{socreview1,socreview2,socreview4,socreview5,socreview6}.

Besides s-wave, the p-wave interacting atomic gases have also attracted great attention in recent years\cite{Chin, thesis}, in particular, in view of the very recent explorations of universal properties  near p-wave resonance\cite{Toronto, Ueda, Yu_p, Zhou}.
%a growing number of p-wave Feshbach resonances achieved in experiments\cite{Chin, thesis}. 
In this work, we study the interplay of SOC and strong p-wave interaction to the short-range and long-range two-body physics. 
%Up to date, the interplay effect between SOC and high partial-wave scattering, such as p-wave, to the basic two-body scattering properties is still unclear, despite a few previous attempts on related problems\cite{}. Giving more and more experimental activities on p-wave Feshbach resonances\cite{..}, it now becomes impetative to look into 
%In view of more and more observations of p-wave Feshbach resonances in atomic gases\cite{..},  
Specifically, we ask the question how would SOC affect the p-wave short-range BC and induce shallow molecules? 
There have been a few related discussions in literature without full answers\cite{Cui, Yu}.  
%has been rarely discussed in literature while a puzzle of it was raised in Ref.\cite{..}.  %been rarely discussed except a few in literature while a puzzle for this has been proposed in Ref.\cite{Cui}. %This question combine two hot research areas in cold Fermi gas, one is synthetic soc and the other is high-partial wave (p-wave) scattering resonance. 
%The p-wave resonance has been observed in a few isotopes of K40 and Li6 fermions. 
Addressing this problem will be fundamentally important for future exploring a new set of few- and many-body physics due to the interplay of SOC and high partial-wave scatterings. 
%in the strongly p-wave interacting system with SOC.

We consider two spin-1/2 fermions subject to an isotropic SOC, which is hopefully realizable in future considering a number of proposals\cite{proposal1,proposal2}. By adopting a two-channel square-well potential generating the p-wave Feshbach resonance\cite{Chin} (see Fig.1a), we can exactly pin down the binding energy of induced molecule as well as its multi-channel structures and short-range behaviors. %extract the short-range asymptotic behavior of molecule wave function and pin down its multi-channel structures as well as the associated binding energy. 
We find that even for the SOC length much longer than the potential range, it can still greatly modify the p-wave short-range BC near p-wave resonances, on contrary to previous findings in the s-wave channel. %One is that the continuity of the wave function and its first derivative at short-range does not guarantee that the p-wave parameter is unchanged; the other is due to the non-perturbative mixing to s-wave channel due to strong p-wave interaction. 
%Such a dramatic response is partly due to the unique sensitivity of p-wave parameters to short-range details, and partly due to the strong interference with s-wave channel induced by SOC. 
Consequently, the conventional p-wave BC cannot predict the induced molecules near p-wave resonance, which instead can be fully destroyed to vanish within a visible range of interaction strength. %shows a visible avoided level crossing. 
Its breakdown is partly due to the intrinsic sensitivity of p-wave scattering parameters to short-range details, and partly due to the strong interference between s- and p-wave channels via SOC. Based on these observations, we finally propose a new p-wave BC, which excellently reproduces the molecule solution in this case and also equally applies for a wide class of single-particle potentials besides SOC.

\begin{figure}[t]
\includegraphics[width=9cm, height=4cm]{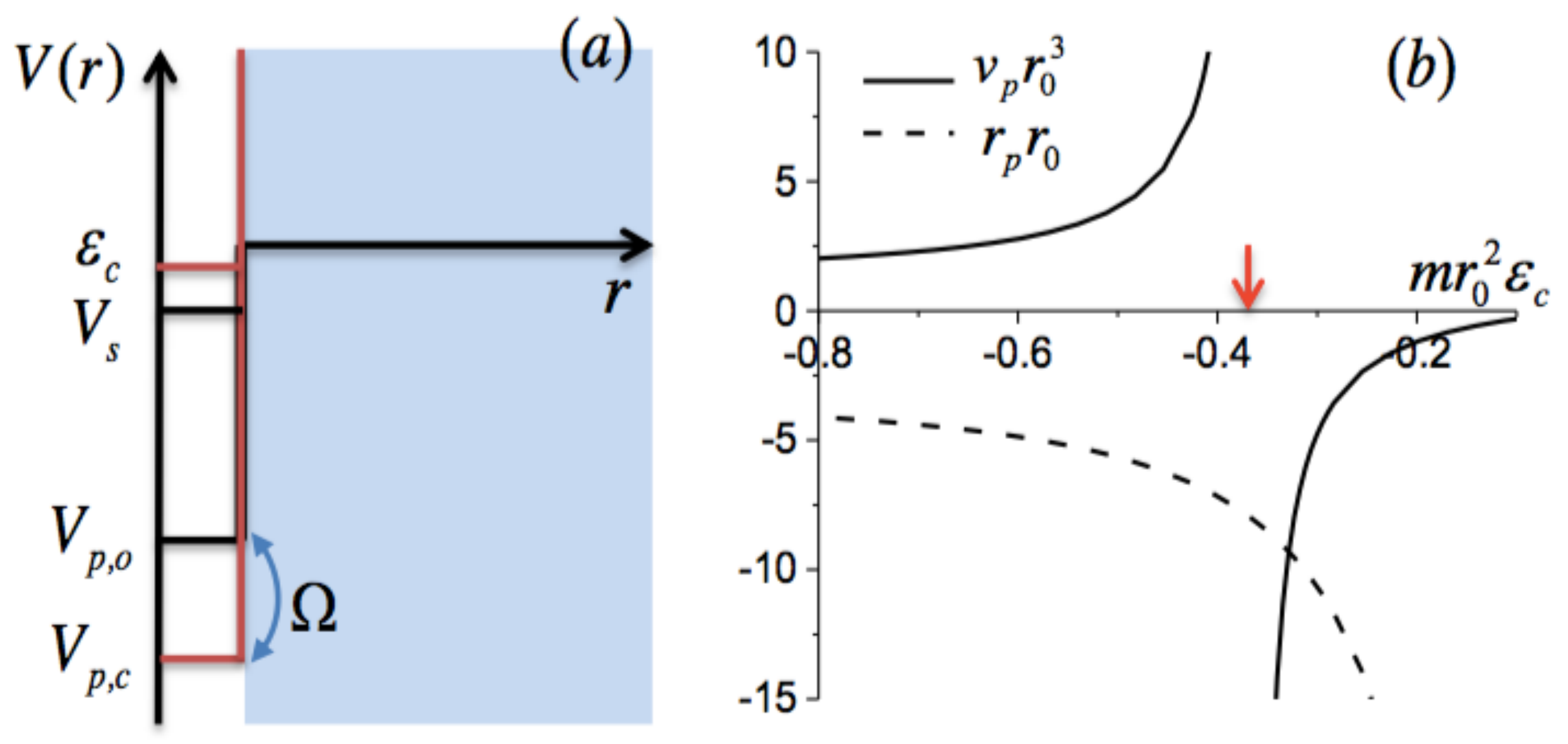}
\caption{(Color online) (a) Schematic plot of s- and p-wave square-well potentials with range $r_0$(see text). %$V_s$ and $V_{p,o/c}$ are respectively the s-wave and open/closed-channel p-wave potential depths at $r<r_0$. $\Omega$ is p-wave inter-channel coupling.
%$V_{p,c}$ supports a bare bound state at $E_c$ near the scattering threshold, resulting in a p-wave resonance. 
The shaded area is the region where single-particle SOC is applied.
(b) p-wave scattering volume ($v_p$) and effective range ($r_p$) as a function of closed-channel molecule $\epsilon_c$ with fixed $v_{bg}=1.2r_0^3$ and $\Omega/|V_{p,c}-V_{p,o}|$ around  $0.12$ across the whole region. The red arrow marks the location of p-wave resonance.
%For the whole plot region we have $\Omega/|V_{p,c}-V_{p,o}|\sim 0.1$.
% Here we use $r_0$ and $E_0=(mr_0^2)^{-1}$ as the units of length and energy.  
} \label{fig1}
\end{figure}

We start from a square-well model potential, as depicted in Fig.1a, for two spin-1/2 fermions with relative distance ${\bf r}$: 
\begin{equation}
V(\cp r)=V_p P_{11;00} + V_s \theta(r_0-r)P_{00;00}. \label{V}
\end{equation}
Here $P_{L_rS;m_lm_s}$ is the projection operator to two-particle state with relative orbital angular momenta $\{L_rm_l\}$ and total spin angular momenta $\{Sm_s\}$. Explicitly, $|Sm_s\rangle$ can be expanded as: $|10\rangle = \frac{|\uparrow\downarrow\rangle+|\downarrow\uparrow\rangle}{\sqrt{2}},  \ |00\rangle = \frac{|\uparrow\downarrow\rangle-|\downarrow\uparrow\rangle}{\sqrt{2}}$. For the p-wave interaction ($L_r=1$), we have selected out $m_l=0$ channel whose resonance can be well separated from the other channels\cite{Chin, thesis}; to mimic the realistic p-wave Feshbach resonance, we adopt a two-channel potential\cite{Chin}  
\begin{equation}
V_p = \left(\begin{array}{cc} V_{p,o} & \Omega \\ \Omega & V_{p,c}\end{array}\right)\theta(r_0-r)+\left(\begin{array}{cc} 0& 0 \\ 0 & \infty \end{array}\right)\theta(r-r_0).
\end{equation}
Here $V_{p,o}$ and $V_{p,c}$ are respectively the open- and closed-channel potentials within the range $r_0$. $V_{p,o}$ provides the background p-wave volume $v_{bg}=-j_2(q_or_0)/(3j_0(q_or_0))$ ($q_o=\sqrt{-mV_{p,o}}$), and $V_{p,c}$ gives a closed-channel molecule at $\epsilon_c$ satisfying $j_1(\sqrt{m(\epsilon_c-V_{p,c})}r_0)=0$. Given $\epsilon_c$ close to the scattering  threshold and an inter-channel coupling $\Omega$, a p-wave resonance can be induced in the open-channel with scattering volume $v_p\rightarrow\infty$ and a finite range $r_p$. Explicitly, $v_p$ and $r_p$ are defined through the phase shift expansion in low-energy limit ($k\ll 1/r_0$) as $k^3\cot\delta_p=-1/v_p+r_pk^2/2$, where the phase shift $\delta_p$ appears in the open-channel wave function $\psi_{p,o}=j_1(kr)-\tan\delta_p n_1(kr)$ and can be determined by requiring $\psi_{p,c}(r_0)=0$ and the continuity of $\psi'_{p,o}/\psi_{p,o}$ at $r_0$ in the current model. In Fig.1b, we plot $v_p$ and $r_p$ as a function of $\epsilon_c$ for a weak $v_{bg}=1.2r_0^3$  and $\Omega\ll |V_{p,c}-V_{p,o}|$. A resonance of $v_p$ occurs at $\epsilon_c=-0.37$ with a finite range $r_p=-7/r_o$.

Importantly, the scattering parameters $v_p$ and $r_p$ also appear in the asymptotic behavior of $\psi_{p,o}\equiv \psi_p$ in the short-range regime $r_0\ll r \ll 1/k$:
\begin{equation}
\psi_{p}(r)\rightarrow \frac{1}{r^2} + \frac{k^2}{2} +\frac{r}{3}(-\frac{1}{v_p}+\frac{r_pk^2}{2})+o(r^3).   \label{short_range}
%\psi_{p,o}(r)\rightarrow (\frac{1}{r^2}-\frac{r}{3v_p}) + k^2(\frac{1}{2} +\frac{r_p r}{2}).
\end{equation}
Thus $v_p, r_p$ determine the ratio between the coefficients of $ r$ and most singular $1/r^2$ terms, which sets the conventional short-range BC in p-wave channel:
\begin{equation}
\frac{(r^2\psi_p)'''}{r^2\psi_p} = %-\frac{r^2}{v_p}+k^2 r\left( 1+\frac{r^3}{v_p} +r_p r (\frac{1}{2}+\frac{r^3}{3v_p})\right) 
2\left( -\frac{1}{v_p}+ \frac{r_pk^2}{2}\right)  +o(r^2), \label{old_BC}
\end{equation}
here the superscript $'''$ denotes the third derivative in terms of $r$. In fact, the pseudo-potential method\cite{LHY, Stock05} and p-wave contact model\cite{Gurarie} all correspond to guaranteeing above short-range BC, regardless of the presence of any external or internal single-particle potentials.
%Explicitly, we have (expression of \delta_p in terms of potential parameters)

Note that Eq.\ref{V} also includes a weak potential $V_s$ in s-wave channel at $r<r_0$, giving the scattering length $a_s/r_0=1-\tan(q_sr_0)/(q_sr_0)$ ($q_s=\sqrt{-mV_s}$) and effective range $r_s$. The reason for its inclusion will be explained later.

Now we consider the SOC part. Different from previous studies\cite{Cui, Yu}, in this work we consider the single-particle SOC applied only to the region outside the interaction potential, as shown by shaded area in Fig.1a, since the laser-generated SOC in experiments\cite{gauge2exp,fermisocexp1,fermisocexp2,2dsoczhang1,Wu2015} can hardly reach the very short-range regime given so deep interaction potentials therein. Thus  SOC will not modify the scattering inside the potential ($r<r_0$). Nevertheless, we will show below that it does greatly modify the p-wave short-range BC in the regime $r_0\ll r\ll 1/k$.

%symmetry analysis; green function

Under the isotropic SOC ($(\lambda/m) \cp{k}\cdot \boldsymbol\sigma$ with $ \boldsymbol\sigma$ the Pauli matrix), the single-particle eigen-state at momentum $\cp k$ has two orthogonal branches $|\cp k^{(\pm)}\rangle$ with eigen-energies 
%\begin{eqnarray}
%|\mathbf{k}^{(+)}\rangle&=&u^{(+)}_{\mathbf{k}}|\mathbf{k}_{\uparrow}\rangle+u^{(-)}_{\mathbf{k}}e^{i\phi_{\mathbf{k}}}|\mathbf{k}_{\downarrow}\rangle,\nonumber\\
%|\mathbf{k}^{(-)}\rangle&=&-u^{(-)}_{\mathbf{k}}e^{-i\phi_{\mathbf{k}}}|\mathbf{k}_{\uparrow}\rangle +u^{(+)}_{\mathbf{k}} |\mathbf{k}_{\downarrow}\rangle;
%\end{eqnarray}
%with $\phi_{\mathbf{k}}=arg(k_x+ik_y)$,
%$u^{(\pm)}_{\mathbf{k}}=\sqrt{\frac{1}{2}\pm \frac{k_z}{2|\mathbf{k}|}}$, and the corresponding eigen-energy
$\epsilon^{(\pm)}_{\cp k}=(k\pm\lambda)^2/(2m)\ (k\equiv|\mathbf{k}|)$. In single-particle level, the isotropic SOC enables the conservation of total angular momentum $\mathbf{j}=\mathbf{l}+\mathbf{s}\ (\mathbf{s}=\frac{1}{2}\boldsymbol\sigma)$. Consequently in two-body level, the total angular momentum $\mathbf{J}=\mathbf{L}+\mathbf{S}$ is also conserved, and the orbital $\mathbf{L}$ can be reduced to the relative component $\mathbf{L}_r$ when we consider the ground state scattering with zero center-of-mass momentum\cite{Cui}.  Given the interaction potential (\ref{V}), we expect the relevant $J$ can be 0 and 2, which can be composed by $S=0,\ L_r=0,2$ or $S=1,\ L_r=1,3$. Such multi-channel structure is revealed clearly in the scattered wave function studied below.

Based on the Lippmann-Schwinger equation, the scattered wave function ($r>r_0$) at energy $E$ reads:
\begin{eqnarray}
\Psi^{sc}(\cp r)=\int d\mathbf{r}' \langle \mathbf{r}|G|\mathbf{r}' \rangle  \langle\mathbf{r}'|V |\Psi\rangle. \label{psi}
\end{eqnarray}
where the Green function can be expanded as
\begin{equation}
\langle\mathbf{r}|G|\mathbf{r}'\rangle=\frac{1}{2}\sum_{\sigma\sigma'=\pm} \frac{\langle \cp r|\cp{k}^{(\sigma)},-\cp{k}^{(\sigma')}\rangle \langle \cp{k}^{(\sigma)},-\cp{k}^{(\sigma')} |\cp r'\rangle}{E-\epsilon_{\cp k}^{(\sigma)}-\epsilon_{-\cp k}^{(\sigma')} + i0^{+}}; \label{G}
\end{equation}
and the interaction part can be parametrized as
\begin{eqnarray}
\langle \mathbf{r}|V|\Psi \rangle=F_p(r) Y_{10}(\Omega_r) |10\rangle+ F_s(r) Y_{00}(\Omega_r) |00\rangle, \ (r<r_0)\label{U_P}
\end{eqnarray}
with $F_s,\ F_p$ respectively the scattering amplitudes in s- and p-wave channel based on Eq.(\ref{V}). 

Lengthy but straightforward calculation of Eq.\ref{psi} indeed results in various $|J;L_rS\rangle$ states. These states include $|2;11\rangle$, $|2;31\rangle$, $|2;20\rangle$,  $|0;11\rangle$, $|0;00\rangle$ generated by p-wave interaction ($F_p$ part in Eq.\ref{U_P}), and $|0;11\rangle$, $|0;00\rangle$ generated by s-wave interaction ($F_s$ part  in Eq.\ref{U_P}). By analyzing their coefficients, it is found that the dominated $1/r^{L_r+1}$ singularity only occur in $L_r=0,1$ channels\cite{note_divergence}, which reads $\psi_p Y_{10}(\Omega_r) |10\rangle + \psi_sY_{00}(\Omega_r) |00\rangle$ with
\begin{equation}
\psi_p=f_p A - f_s B;\ \ \ \ \psi_s= f_s C+f_p D. \label{psi_sp}
\end{equation}
Here $f_s=\int_0^{r_0} dr r^2F_s(r)$, $f_p=\int_0^{r_0} dr r^3F_s(r)$, and $A,B,C,D$ are the functions of $r$ and $E^{+}\equiv E+i0^+$:%\cite{supple}: 
\begin{eqnarray}
A&=&\frac{2}{3\pi} \int dk k^3 j_1(kr)\left( \sum_{\sigma=\pm} \frac{3/10}{E^+-2\epsilon_{\cp k}^{(\sigma)}} + \frac{2/5}{E^+-\epsilon_{\cp k}^{(+)}-\epsilon_{-{\cp k}}^{(-)}}\right); \nonumber\\
B&=&\frac{i}{\sqrt{3}\pi} \int dk k^2 j_1(kr)\left( \frac{1}{E^+-2\epsilon_{\cp k}^{(+)}} - \frac{1}{E^+-2\epsilon_{\cp k}^{(-)}}\right); \nonumber\\
C&=&\frac{1}{\pi} \int dk k^2 j_0(kr)\left( \frac{1}{E^+-2\epsilon_{\cp k}^{(+)}} + \frac{1}{E^+-2\epsilon_{\cp k}^{(-)}}\right); \nonumber\\
D&=&\frac{i}{3\sqrt{3}\pi} \int dk k^3 j_0(kr)\left( \frac{1}{E^+-2\epsilon_{\cp k}^{(+)}} - \frac{1}{E^+-2\epsilon_{\cp k}^{(-)}}\right). \nonumber
\end{eqnarray}
Note that we have used $\epsilon_{\cp k}^{(\sigma)}=\epsilon_{-\cp k}^{(\sigma)}$ to simplify above equations. In small $r$ limit, $A\sim 1/r^2$, and $C, D\sim 1/r$. Here $D\sim 1/r$  shows an interesting fact that the p-wave interaction alone ($f_p\neq0,\ f_s=0$) can cause short-range singularity in the s-wave channel\cite{Cui}. Thus to ensure a physical $\psi_s$ near $r=0$, an s-wave potential has to be activated at short-range, as shown in Fig.1a. In other words, one has to explicitly include s-wave interaction while considering the physics  near p-wave resonance. Note that this is in sharp difference to the pure s-wave interaction ($f_p=0,\ f_s\neq0$), which does not induce any singularity in p-wave channel\cite{Cui, Vijay} and a single s-wave model is adequate to describe the physics near s-wave resonance.

Next we study the bound state solution with $E=-\kappa^2/m$, whose wave function is fully given by $\Psi^{sc}$ (Eq.\ref{psi}) for $r>r_0$. Specifically, $\kappa$ and the ratio $f_{sp}\equiv-if_s/f_p$ can be exactly determined by requiring the continuity of both $\psi_s'/\psi_s$ and $\psi_p'/\psi_p$ at the potential boundary $r=r_0$. 
%Next we study the bound state solution, with the wave function given by $\Psi^{sc}$ (Eq.\ref{psi}) for $r>r_0$. Since the interaction potential (\ref{V}) only applies in $L_r=S=0$ (s-wave)  and $L_r=S=1$(p-wave) channels, we can solve the energy $E=-\kappa^2/m$ and ratio $f_{sp}=-if_s/f_p$ by requiring the continuity of $\psi'/\psi$ at $r=r_0$ individually for these two channels. 
These quantities also determine the asymptotic  behaviors of $\psi_s$ and $\psi_p$ (Eq.\ref{psi_sp}) in short-range regime ($r_0\ll r \ll 1/\kappa$):
\begin{eqnarray}
\psi_p &\rightarrow& \frac{1}{r^2} +(-\frac{\kappa^2}{2}-\sqrt{3}f_{sp}\lambda+0.7\lambda^2)+\eta_p r; \label{short_psip}\\
\eta_p &=& \frac{f_{sp}}{\sqrt{3}\kappa}\lambda(3\kappa^2-\lambda^2)+\frac{\lambda^4-6\lambda^2\kappa^2+\kappa^4}{5\kappa} +\frac{2(\lambda^2+\kappa^2)^{3/2}}{15}, \nonumber\\
\psi_s &\rightarrow& \frac{1}{r}+\eta_s; \\
\eta_s&=&\frac{3\sqrt{3}f_{sp}(\lambda^2-\kappa^2)-\lambda(\lambda^2-3\kappa^2)}{(3\sqrt{3}f_{sp}-2\lambda)\kappa}. \nonumber
\end{eqnarray}
%and $\psi_s \rightarrow 1/r+\eta_s$, with
%\begin{equation}
%\eta_s=\frac{3\sqrt{3}f_{sp}(\lambda^2-\kappa^2)-\lambda(\lambda^2-3\kappa^2)}{(3\sqrt{3}f_{sp}-2\lambda)\kappa}
%\end{equation}
One can check that in $\lambda=0$ limit, these equations well reproduce the free-space results $\eta_s=-\kappa$ and $\eta_p=\kappa^3/3$. When turn on SOC ($\lambda\neq 0$), under certain limit of $f_{sp}$ they can reproduce the results from individual s- or p-wave BC. 
% (equivalent to using individual s/p-wave pseudo-potentials\cite{LHY, Stock05} or contact models\cite{review_Fermi_gas, Gurarie}). 
Namely, the individual s-wave BC corresponds to equating $\eta_s$ with $1/a_s-r_s\kappa^2$ in $f_{sp}\rightarrow\infty$ limit\cite{Vijay, Cui}; while the individual p-wave BC corresponds to equating $\eta_p$ with $-\frac{1}{3}(\frac{1}{v_p}+\frac{r_p}{2}\kappa^2)$ in $f_{sp}\rightarrow0$ limit\cite{Cui}. Of course  one can further improve the theory by relaxing $f_{sp}$ and imposing the s- and p-wave BC simultaneously. We will show below that none of these theories predict correctly the bound states near p-wave resonance. 

\begin{figure}[t]
\includegraphics[width=8.3cm]{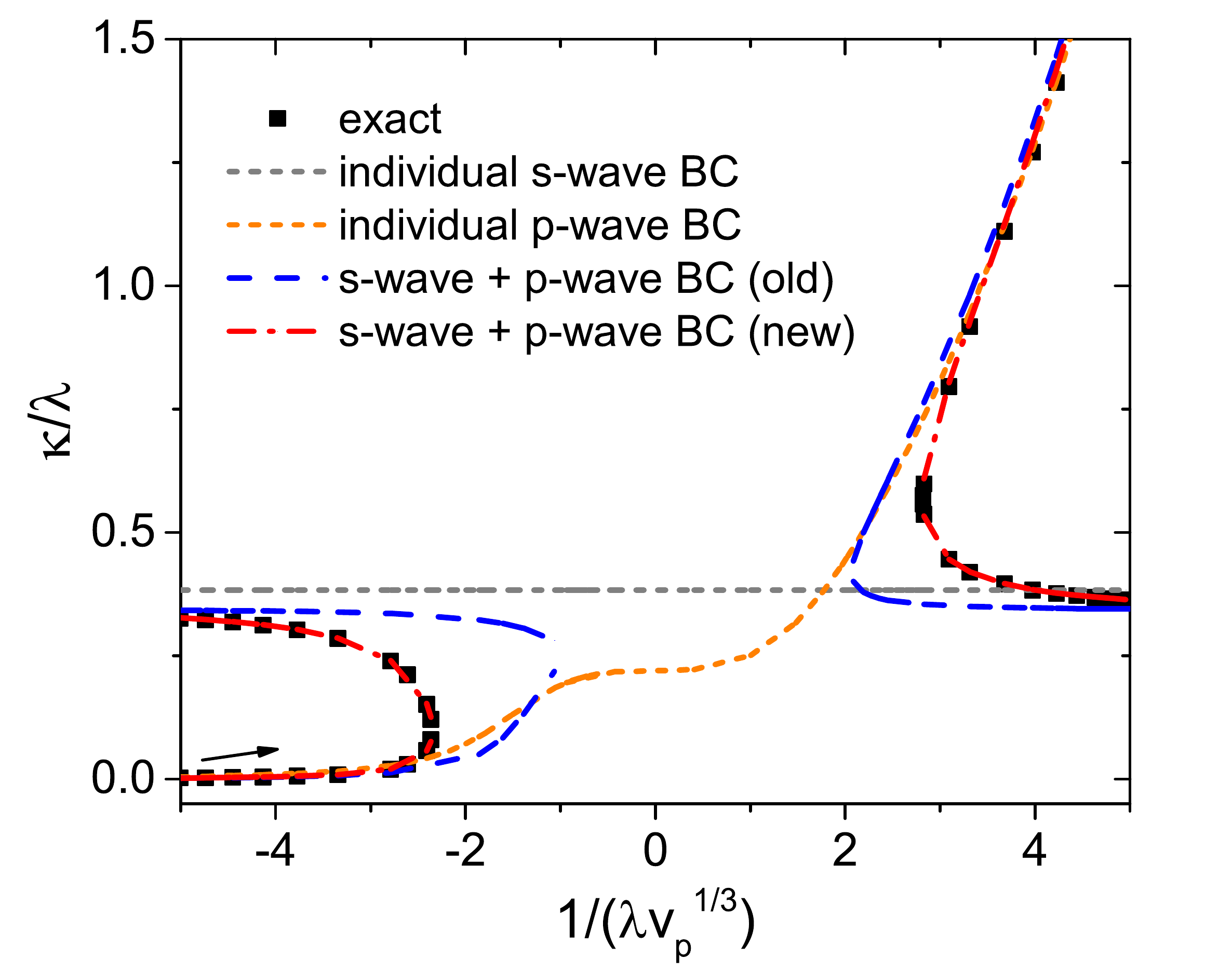}
\caption{(Color online) Bound state solution across p-wave resonance.  $\lambda r_0=0.1$, $a_s=-4.5 r_0$. Square dots show exact solutions by solving the square-well potential in Fig.1a (i.e., by requiring the continuity of $\psi_s'/\psi_s$ and $\psi_p'/\psi_p$ at $r=r_0$). For comparison, we also show results  from individual s/p-wave BC (gray/orange short-dashed lines), from the combined s-wave and conventional p-wave BC (blue dashed; by requiring the continuity of $\psi_s'/\psi_s$ at $r_0$ and matching $\eta_p$ to $-\frac{1}{3}(\frac{1}{v_p}+\frac{r_p}{2}\kappa^2)$, see Eq.\ref{old_BC}), and from the combined s-wave and new p-wave BC (red dashed-dot; by requiring the continuity of $\psi_s'/\psi_s$ at $r_0$ and matching $\psi_p$ in Eq.\ref{short_psip} to Eq.\ref{new} at $r_0$). } \label{fig2}
\end{figure}

In Fig.2, we plot the exact solution of $\kappa/\lambda$ across the p-wave resonance (square dots), taking a small SOC strength $\lambda r_0=0.1$ and a weak s-wave interaction with $a_s=-4.5 r_0$. It shows two branches of solutions. Far away from p-wave resonance, the two branches follow the predictions from individual s-wave and individual p-wave BC(short-dashed lines), respectively giving the s-wave and p-wave dominated molecules. However, when close to p-wave resonance with $1/(\lambda v_p^{1/3})\sim[-3, 3]$, 
%the two branches show a visible avoided crossing, suggesting a strong interference between s- and p-wave channels. In this regime, both the individual BC (short-dashed) and the combined s- and conventional p-wave BC (blue dashed) fail to give the correct answer. 
the exact solutions no longer follow the predictions from individual s-wave or individual p-wave BC, nor from the combined s-wave and conventional p-wave BC (Eq.\ref{old_BC}) (blue dashed). Instead, no molecule solution is found in this regime, which can be attributed to the enhanced interference between s- and p-wave channels due to the presence of SOC and strong p-wave interaction.  

%Remarkably, such an interference will invalidate the usual p-wave short-range boundary condition even near p-wave resonance, as we can see from the clear deviation between exact solutions and results from the improved theory with relaxed $f_{sp}$ and the usual p-wave boundary condition (dot-dashed line). 

\begin{figure}[h]
\includegraphics[width=7.5cm]{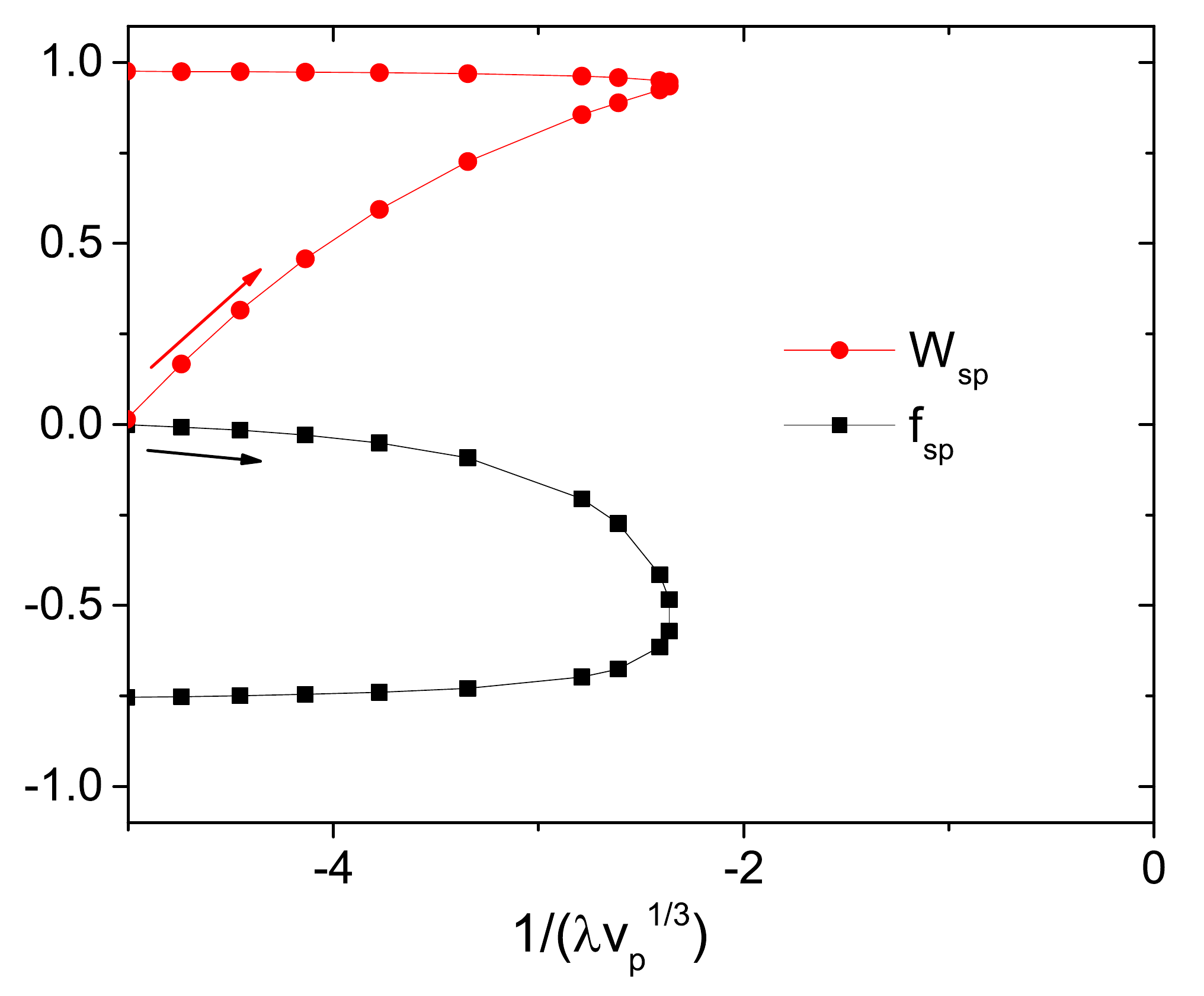}
\caption{(Color online) $f_{sp}$ and its associated weight $W_{sp}$ in $\eta_p$ for one branch of molecule solution in Fig.2. The black and red arrows show the change of $f_{sp}$ and $W_{sp}$ following the trajectory also arrow-marked in Fig.2.
} \label{fig3}
\end{figure}

In the following, we show that the SOC-induced s-p interference (resulting in a finite $f_{sp}$) greatly modify the p-wave asymptotic parameter $\eta_p$, such that the usual p-wave BC breaks down even near the p-wave resonance. To see this clearly, in Fig.3 we plot $f_{sp}$ and its associated weight in $\eta_p$ ($W_{sp}$, i.e., the weight of the first term in $\eta_p$) for  one branch of solution shown in Fig.2. We see that as $v_p$ is tuned to approach resonance,  $|f_{sp}|$ gradually increases from zero to around $0.5$, and $W_{sp}$ increases from zero to nearly 1, suggesting the molecule evolves from the p-wave dominated to s-p strongly interfered regime. In the latter regime, the s-p interference play an essential role in determining the actual $\eta_p$, which can be very different from the predictions of usual p-wave BC (Eq.\ref{short_range}). 
%and the p-wave short-range boundary condition has to be rectified. 

In comparison to the robust s-wave BC near s-wave resonance under SOC\cite{Cui,Yu,Zhang}, here the fragile p-wave BC near the p-wave resonance has its intrinsic reasons. This is because the %usual p-wave short-range behavior is characterized by the ratio between very singular $1/r^2$ and very weak $r$ terms as shown in Eq.\ref{short_range}. Such ratio can be very easily destroyed by perturbations outside the potential, since it is related to the third derivative of $\psi_p$ 
usual p-wave BC (Eq.\ref{old_BC}) is characterized by the ratio between very singular $1/r^2$ and very weak $r$ terms as shown in Eq.\ref{short_range}. Such ratio can be very easily destroyed by perturbations outside the potential, since it is related to the third derivative of $\psi_p$ (see Eq.\ref{old_BC}) while any realistic BC can only guarantee the continuity of $\psi_p$ and its first derivative across the boundary. This is quite contrary to the s-wave case where the short-range behavior is characterized by the ratio between $1/r$ and constant terms. In this case 
%the continuity of $\psi_s$ and its first derivatives at the potential boundary exactly guarantee that 
the continuity of $\psi_s$ and its first derivative is adequate to guarantee the continuity of this ratio. Therefore the short-range behavior of $\psi_s$ outside the potential ($r>r_0$) is universally given by the physics inside the potential ($r<r_0$), which is hardly modified by small perturbations such as SOC.

Motivated by above observations, we can set up a new p-wave BC following the persistent continuity of $\psi_p'/\psi_p$ at the short-range boundary. Given the wave function inside the potential is not modified by SOC, the new BC can be formulated using the original scattering parameters $v_p$ and $r_p$ in free space (see Eq.\ref{short_range}). Namely, at the potential boundary where SOC just starts to take effect, we have ($r\ll k^{-1},\ |v_p|^{1/3}$) 
\begin{equation}
\frac{(r^2\psi_p)'}{r^2\psi_p} = %-\frac{r^2}{v_p}+k^2 r\left( 1+\frac{r^3}{v_p} +r_p r (\frac{1}{2}+\frac{r^3}{3v_p})\right) 
\left( -\frac{1}{v_p}+ \frac{r_pk^2}{2}\right) r^2 +k^2 r +o(r^4). \label{new}
\end{equation}
Note that this rectified BC reflects the whole structure of $\psi_p$ in short-range limit (Eq.\ref{short_range}), including the constant term which has been omitted by usual p-wave BC (Eq.\ref{old_BC}). Because this constant is more dominated than $r$ term in short-range limit, and because the SOC greatly modifies such constant (Eq.\ref{short_psip}), it is important for its effect to be taken into account. In Fig.2, we plot the re-calculated $\kappa$ (red dashed-dot) using Eq.\ref{new} to replace the usual p-wave BC for $\psi_p$ at $r=r_0$. The obtained results are excellently consistent with the exact solutions across p-wave resonance. We also check its robustness by choosing other boundaries $r(>r_0)$ in matching Eq.\ref{new}, and find good consistence with exact solutions for $r$ up to a few times of $r_0$. This justifies Eq.\ref{new} as the correct p-wave BC in the presence of SOC. In fact, since Eq.\ref{new} only relies on the wave function continuity near the potential boundary, it will be generally applicable for any type of SOC or other single-particle potentials, as long as they do not modify the physics inside the short-range potential.

In summary, we have studied the interplay of an isotropic SOC and strong p-wave interaction to both the short-range physics and shallow molecules of two interacting fermions. We find that even for SOC length much longer than the range of interaction potential, it can still induce strong interference between s- and p-wave channels, which leads to the vanishing of molecules near p-wave resonance and 
%a visible avoided crossing between molecule branches as well as 
the breakdown of usual p-wave short-range BC. The proposed new p-wave BC, which applies for a wide class of single-particle potentials including SOC, will hopefully play more roles in dealing with few- and many-body problems near p-wave resonance. 

Finally, since our scheme to solve the two-body problem and the conclusion of s-p interference do not rely on the specific type of SOC, 
%the avoided crossing of molecule branches may even be observable in 
our results will shed light on the molecule formation in current experiments with one or two-dimensional SOC\cite{gauge2exp,fermisocexp1,fermisocexp2, 2dsoczhang1, Wu2015} and near p-wave resonance. In particular, it would be practical to realize SOC in quite a number of two-species fermion systems with p-wave Feshbach resonance, such as the $F=9/2$ $^{40}$K Fermi gas with two hyperfine states $|m_{F1}=-9/2; m_{F2}=-5/2\rangle$ at $B_0=215$G, $|-3/2;-1/2\rangle$ at $338$G, and $|9/2;5/2\rangle$ at $44$G\cite{thesis}. It is interesting to explore in future the multichannel molecules in these systems.%We expect such avoided crossing most notable near p-wave resonance where the p-wave molecules strongly hybridize with s-wave ones in the presence of a background s-wave interaction. 

%realization:
%a+c(-9/2+-5/2, in ml=0 channel) p-wave resonance of K40 at B0=215G --- Munich expe
%d+e (-3/2+-1/2, in ml=0 channel) ... 338G ---  Amsterdam
%h+j (5/2+9/2, in ml=0 channel) ... 44G --- Amsterdam

{\bf Acknowledgment.}  I like to thank Doerte Blume, Qingze Guan and Peng Zhang for useful discussions on the few-body physics with spin-orbit coupling during the KITP program "Universality in Few-Body Systems" in the winter of 2016. This project is supported by the National Natural Science Foundation of China (No.11374177, No.11626436, No. 11421092, No. 11534014), and the National Key Research and Development Program of China (2016YFA0300603).


\begin{thebibliography}{}

\bibitem{review_1}M. Z. Hasan and C. L. Kane, Rev. Mod. Phys. {\bf 82}, 3045 (2010).

\bibitem{review_2}X.-L. Qi and S.-C. Zhang, Rev. Mod. Phys. {\bf 83}, 1057 (2011).

\bibitem{gauge2exp} Y. J. Lin, K. Jim\'{e}nez-Garc\'{i}a, and I. B. Spielman Nature, {\bf 471}, 83 (2011).

\bibitem{fermisocexp1} P. Wang, Z. Q. Yu, Z. Fu, J. Miao, L. Huang, S. Chai, H. Zhai, and J. Zhang, Phys. Rev. Lett. {\bf 109}, 095301 (2012).

\bibitem{fermisocexp2} L. W. Cheuk, A. T. Sommer, Z. Hadzibabic, T. Yefsah, W. S. Bakr, and M. W. Zwierlein, Phys. Rev. Lett. {\bf 109}, 095302 (2012).

\bibitem{2dsoczhang1} L. Huang, Z. Meng, P. Wang, P. Peng, S. Zhang, L. Chen, D. Li, Q. Zhou, and J. Zhang, Nature Phys. {\bf 12}, 540 (2016).

\bibitem{Wu2015} Z. Wu, L. Zhang, W. Sun, X.-T. Xu, B.-Z. Wang, S.-C. Ji, Y. Deng, S. Chen, X.-J. Liu, and J.-W. Pan, Science {\bf 354}, 83 (2016).


% FR:
\bibitem{Chin} C. Chin, R. Grimm, P. Julienne and E. Tiesinga, Rev. Mod. Phys. {\bf 82}, 1225 (2010).

%pseudopotential:
\bibitem{LHY}K. Huang and C. N. Yang, Phys. Rev. {\bf 105}, 767 (1957). 
\bibitem{Stock05}R. Stock, A. Silberfarb, E. L. Bolda and I. H. Deutsch, Phys. Rev. Lett. {\bf 94}, 023202 (2005).

\bibitem{review_Fermi_gas} S. Giorgini, L. P. Pitaevskii and S. Stringari, Rev. Mod. Whys. {\bf 80}, 1215 (2008).

\bibitem{Gurarie}V. Gurarie, L. Radzihovsky, Annual of Physics {\bf 322}, 2 (2007).

%for s-wave case, soc only take perturbative effect on modification of short-range physics:
\bibitem{Cui}X. Cui, Phys. Rev. A {\bf 85}, 022705 (2012).
\bibitem{Yu}Z. Yu, Phys. Rev. A {\bf 85}, 042711 (2012). Y. Wu and Z. Yu, Phys. Rev. A {\bf 87}, 032703 (2013).
\bibitem{Zhang}P. Zhang, L. Zhang, and Y. Deng, Phys. Rev. A {\bf 86}, 053608 (2012); L. Zhang, Y. Deng and P. Zhang, Phys. Rev. A {\bf 87}, 053626 (2013).

%for all energies (lowest energy manifold):
\bibitem{Blume}Q. Guan, D. Blume, Phys. Rev. A {\bf 94}, 022706 (2016)

%scattering formulism for positive energy:
\bibitem{You}H. Duan, L. You, and B. Gao,  Phys. Rev. A {\bf 87}, 052708 (2013).
\bibitem{Greene}S.-J. Wang and C. H. Greene,  Phys. Rev. A {\bf 91}, 022706 (2015).

%2-body:
\bibitem{Vijay}J. P. Vyasanakere and V. B. Shenoy, Phys. Rev. B {\bf 83} 094515
(2011); {\it ibid}, New. J. Phys {\bf 14}, 043041 (2012).

%3-body:
\bibitem{trimer_soc} Z.-Y. Shi, X. Cui, and H. Zhai, Phys. Rev. Lett. {\bf 112}, 013201 (2014); Z.-Y. Shi, H. Zhai, and X. Cui, Phys. Rev. A {\bf 91}, 023618 (2015).
\bibitem{borromean_soc} X. Cui and W. Yi, Phys. Rev. X {\bf 4}, 031026 (2014).


\bibitem{socreview1} V. Galitski and I. B. Spielman, Nature (London) {\bf 494}, 49 (2013).

\bibitem{socreview2} N. Goldman, G. Juzeli\={u}nas, and P. \"Ohberg, I. B. Spielman, Rep. Prog. Phys. {\bf 77}, 126401 (2014).


\bibitem{socreview4} H. Zhai, Rep. Prog. Phys. {\bf78}, 026001 (2015).

\bibitem{socreview5} W. Yi, W. Zhang, and X. Cui, Sci. China: Phys. Mech. Astron. {\bf 58}, 014201 (2015).

\bibitem{socreview6} J. Zhang, H. Hu, X. J. Liu, and H. Pu, Ann. Rev. Cold At. Mol. {\bf 2}, 81 (2015).

%thesis:
\bibitem{thesis}PhD thesis of A. Ludewig, ``Feshbach resonance in $^{40}$K", http://hdl.handle.net/11245/1.369033

%p-wave contact:
\bibitem{Toronto}C. Luciuk, S. Trotzky, S. Smale, Z. Yu, S. Zhang, J. H. Thywissen, Nature Physics {\bf 12}, 599 (2016).
\bibitem{Ueda} S. M. Yoshida, M. Ueda, Phys. Rev. Lett. {\bf 115}, 135303 (2015).
\bibitem{Yu_p}Z. Yu, J. H. Thywissen, S. Zhang, Phys. Rev. Lett. {\bf 115}, 135304 (2015); see also Erratum: Phys. Rev. Lett. {\bf 117}, 019901 (2016).
\bibitem{Zhou}M.-Y. He, S.-L. Zhang, H. M. Chan, Q. Zhou,  Phys. Rev. Lett. {\bf 116}, 045301 (2016).


% 3d soc proposal:
\bibitem{proposal1}B. M. Anderson, G. Juzeliunas, V. M. Galitski, and I. B. Spielman, Phys. Rev. Lett. {\bf 108}, 235301 (2012).
\bibitem{proposal2}B. M. Anderson, I. B. Spielman, and G. Juzeliunas,  Phys. Rev. Lett. {\bf 111}, 125301 (2013).


\bibitem{note_divergence} We also find a {\it weak} singularity as $ \lambda/r$ in $|2;20\rangle$ state, while the dominated singularity $1/r^3$ is absent. Therefore we consider it a negligible short-range consequence to d-wave channel and discard it in this work. 

%\bibitem{supple} See supplementary material for more details.




%\bibitem{E-dependence}D. Blume and C. H. Greene, Phys. Rev. A, {\bf 65}, 043613(2002); E. L. Bolda, E. Tiesinga and P. S. Julienne, Phys. Rev. A {\bf 66}, 013403 (2002),


%\bibitem{Gurarie}V. Gurarie and L. Radzihovsky, Annuals of Physics, {\bf 322}, 2 (2007).

%\bibitem{Yip}S.-K. Yip, Phys. Rev. A {\bf 78}, 013612 (2008); S.-G. Peng, S.-Q. Li, P. D. Drummond and X.-J. Liu, arxiv:1107.2740.


\end{thebibliography}
\end{document}